\documentclass[runningheads]{llncs}
\usepackage[T1]{fontenc}
\usepackage{graphicx}
\usepackage{subcaption}
\usepackage{hyperref}
\usepackage{color}

\usepackage[range-phrase=--]{siunitx}

\usepackage[table]{xcolor} 

\usepackage[english]{babel}
\usepackage[autostyle]{csquotes}

\usepackage{algorithm}
\usepackage{algorithmicx}
\usepackage{algpseudocode}
\usepackage{amsmath,amsfonts}

\usepackage{booktabs}
\usepackage{multirow}
\usepackage{makecell}

\usepackage{xcolor} 
\usepackage[inline]{enumitem}
\usepackage{comment}

\newcommand{\NA}{{\color{gray}N/A}}

\usepackage{chngcntr}

\usepackage[inkscapelatex=false]{svg} 

\begin{document}
\title{FedADAS: Communication-Efficient Federated Distillation for On-Device Driver Yawn Recognition in Vehicular Networks
\thanks{This paper is accepted in the 28th International Conference on Pattern Recognition (ICPR) 2026.}
}
\titlerunning{FedADAS}

\author{Ahmed Mujtaba\inst{1} \and
Gleb Radchenko\inst{1} \and
Marc Masana\inst{2} \and
Radu Prodan\inst{3}}
\authorrunning{A. Mujtaba et al.}
\institute{Embedded Systems Division, Silicon Austria Labs, Graz, Austria \and
Institute of Visual Computing, Graz University of Technology, Austria \and
Department of Computer Science, University of Innsbruck, Austria
\email{\{ahmed.mujtaba, gleb.radchenko\}@silicon-austria.com} \\
\email{mmasana@tugraz.at} \qquad
\email{radu.prodan@uibk.ac.at}
}
\maketitle              

\begin{abstract}
Driver fatigue is a critical safety concern in advanced driver assistance systems. Driver monitoring models trained off-site on static datasets adapt poorly to real-world conditions, while standard federated learning imposes high communication overhead, assumes homogeneous architectures, and struggles with personalized driver data. We present FedADAS, a federated distillation framework enabling collaborative on-device learning across heterogeneous vehicular networks. FedADAS enables full model heterogeneity by exchanging only soft logits on a shared public dataset, allowing each vehicle to run a customized model tailored to its computational constraints. Additionally, we introduce a yawn recognition pipeline supporting training and inference on edge devices that provides two robust architectures: Performance-Efficient (\num{99.7}~MB) achieving \num{98.3}\% F1-score with \num{1.99}ms inference time on a Jetson NANO, and a Memory-Efficient (\num{0.6}~MB) that trains an epoch in \num{6.12} minutes on a Jetson AGX Orin. In experiments with up to 115 edge clients, FedADAS significantly outperforms traditional federated learning approaches at higher client participation, achieving up to \num{9974}$\times$ reduction in communication cost while maintaining a superior tradeoff between personalization and generalization under extreme data heterogeneity, demonstrating its suitability for real-world deployment. Code is available at \url{https://opensource.silicon-austria.com/mujtabaa/fedadas}.

\end{abstract}
\section{Introduction}
Computer vision, powered with deep learning (DL) models, plays a crucial role in advanced driver assistance systems by interpreting conditions in and around the vehicle and triggering safety interventions. In-cabin monitoring systems and driver monitoring systems (DMS) are key ADAS applications that assess the driver and passenger's state and behavior to ensure safety and improve the driving experience~\cite{dms}. Driver fatigue detection is a core DMS feature that helps prevent road accidents since fatigue significantly impairs a driver’s alertness and reaction time~\cite{nthsa}. DL–based perception models are widely employed in DMS to detect indicators of driver fatigue, including facial cues, head pose, gaze direction, and eye opening and closure patterns~\cite{dms}.

To enable in-vehicle decisions and ensure low-latency responses, DMS perception models must adapt to the energy and performance constraints of the edge computing nodes installed in vehicles, reducing reliance on cloud platforms. Recent DMS research emphasizes optimized DL models to enable real-time, low-latency on-board inference, reducing dependence on cloud connectivity and supporting continuous in-vehicle operation~\cite{dms}. However, these models are still trained off-site on static datasets, limiting their adaptability to the diverse conditions encountered in real-world driving environments. Repeatedly fine-tuning the models to real-driving environments through on-device learning also entails the risk of overfitting on local data.

Federated learning (FL)~\cite{mcmahan2017communication} enables edge nodes to collaboratively train generalized DL models on locally available data by exchanging model parameters with a central server, which aggregates and redistributes updated models globally, allowing their continuous improvement across the fleet. However, several critical obstacles in edge-based vehicular settings prevent the adoption of FL.
\begin{enumerate}[itemsep=0.3cm]
    \item \textit{Device heterogeneity}, unable to host a common DL model architecture compatible with FL, or incurring unacceptably high local training time.
    \item \textit{Communication overhead} for frequent exchange of model parameters, critical on vehicles with limited or low-power hardware connectivity~\cite{lan2023communication}.
    \item \textit{Non-independent and identically distributed (non-IID) local data} due to driver-specific behavior, sensor performance, and environment conditions, complicating global model convergence.
\end{enumerate}
\emph{Federated distillation (FD)}~\cite{FedMD} addresses the challenges of FL by eliminating the requirement of sharing a common DL model architecture, enabling edge devices to run a customized model adapted to their resource capabilities. To transfer knowledge, FD clients share only model outputs (\textit{logits}) on a common public dataset, significantly reducing communication overhead~\cite{mujtaba2025federated}. This works makes the following technical contributions:

\begin{itemize}
    \item We present \emph{FedADAS}, a first FD-based framework for collaborative learning among vehicular networks with complete model heterogeneity. FedADAS enables vehicles to train local DL models collaboratively by sharing \emph{logits} on a shared public dataset with the server for aggregation. The aggregated logits on the server act as an ensemble knowledge for the vehicles in distilled form. A scalability test with over 100 vehicles for a yawn classification task establishes that FedADAS significantly outperforms traditional FL~\cite{mcmahan2017communication} with nearly \num{9974}$\times$ communication cost reduction, and provides the best personalization and generalization tradeoff for local DL models in cross-vehicular learning scenarios.
    \item We evaluate FedADAS on driver yawn recognition as a representative DMS classification task; the underlying logit-exchange and KD mechanisms make no task-specific assumptions and are applicable to multi-class perception tasks. Additionally, we introduce a lightweight DMS pipeline for yawn recognition that meets the constraints of edge computing platforms and incorporates two robust yawn classification architectures: Memory-Efficient and Performance-Efficient. We comprehensively evaluate the DMS pipeline using YawDD+~\cite{yawdd+} on NVIDIA Jetson AGX and NANO. 
\end{itemize}

\begin{table}[t]
\centering
\caption{Related work summary.}
\label{tab:related_works}
\resizebox{\textwidth}{!}{%
\begin{tabular}{@{}lcccccc@{}}
\toprule
\textit{Related work} & \textit{Method} & \textit{Model heterogen.} & \textit{Public data} & \textit{Clients} & \textit{Edge train} & \textit{Shared information} \\ \midrule
FedHotpot~\cite{wang2024federated} & FL & No & No & 20 & No & model params \\ 
pFedVs~\cite{liao2025personalized} & FL & No & No & 20-100 & No & model params \\ 
FedLane~\cite{eid2024federated} & FL & No & No & 3 & No & model params \\ 
FedBiKD~\cite{shang2023fedbikd} & FL and KD & No & Yes & 20 & No & model params \\ 
FedCMD~\cite{bano2024fedcmd} & hybrid FL-KD & No & Yes & 4-30 & No & model params; soft logits \\ 
Data-Free~\cite{data-free} & hybrid FL-KD & No & No & 3--10 & No & model params; interm. features \\ 
FedDLD~\cite{xiao2025feddld} & hybrid FL-KD & No & No & 20--50 & No & model params; soft logits \\ 
DB-EPFD~\cite{huang2025environment} & hybrid FL-KD & Partial & No & 10--20 & No & partial model params \\ 
FedADAS (Ours) & FD & Yes & Yes & 3--115 & Yes & soft logits \\ \bottomrule
\end{tabular}}
\end{table}

\section{Related Works}
\label{sec:related_works}
\emph{Collaborative learning} has emerged as a promising approach among vehicles for ADAS~\cite{NVIDIA_FL_AV}. Existing works summarised in Table~\ref{tab:related_works} employ FL and knowledge distillation (KD) to transfer knowledge among vehicles for different ADAS applications. These works share either:
\begin{enumerate*}
    \item \emph{model parameters} trained on vehicle local data (used in traditional FL frameworks FedHotpot~\cite{wang2024federated}, pFedVs~\cite{liao2025personalized}, FedLane~\cite{eid2024federated}) requiring all vehicles to share same model architecture for accurate aggregation on server; 
    \item \emph{intermediate features} of vehicle local model with identical shapes for accurate aggregation;
    \item \emph{soft logits} produced by vehicle local model on shared unlabeled public data for KD.
\end{enumerate*}
Hybrid FL-KD frameworks, such as FedBiKD~\cite{shang2023fedbikd}, FedCMD~\cite{bano2024fedcmd}, Data-Free~\cite{data-free}, and FedDLD~\cite{xiao2025feddld}, incorporate KD but still require homogeneous DL architectures across clients due to the sharing of model parameters. DB-EPFD~\cite{huang2025environment} achieves partial model heterogeneity by allowing different decision heads while sharing the same model backbone parameters. None of the prior methods support full model heterogeneity, since they are not designed to operate when each client runs a structurally different architecture. FedADAS is the first FD-based framework to achieve complete model heterogeneity, on-device training on edge hardware, and evaluation at fleet scale up to 115 heterogeneous clients. The contribution is therefore a first validated systems-level realization of FD for vehicular edge AI.

\emph{Video-based methods} for DMS yawn recognition sample consecutive frames from full-length videos to serve as a single input volume for training. Existing studies attaining over \qty{93}{\percent} accuracy include a two-stream spatial–temporal graph convolutional network (2s-STGCN) \cite{bai2021two}, a hybrid convolutional-recurrent neural network (CNN-RNN) model to incorporate spatial-temporal features during training \cite{majeed2023detection}, and a dual-lightweight swin transformer model \cite{xu2025novel}. However, these video-based methods demand considerable computational resources making them impractical to train on resource-constrained edge devices~\cite{wang2025empowering}, 

\emph{Frame-based methods} are more lightweight, extract individual frames from videos, and are capable of achieving similar accuracy above \qty{93}{\percent}. Existing studies encompass: a CNN-based driver fatigue system with an overall performance of six frames per second (FPS) on NVIDIA Jetson NANO \cite{civik2023real}, a distracted driving detection system based on improved YOLO8 with 43.17 FPS on NVIDIA Jetson Xavier NX~\cite{al2025real}, a driver fatigue monitoring system based on MobileNetv3 with 22 FPS on NVIDIA Jetson TX2~\cite{zhou2021real}, and a two-staged CNN with 96.3ms inference time on Raspberry Pi 4~\cite{he2020real}.
Although these studies evaluate model inference on edge devices, they do not examine whether training can also be performed on such devices, a capability that is vital for generalized and adaptive DMS and FL solutions. This study closes the gap of training DL architectures on edge devices by introducing two (Memory-Efficient and Performance-Efficient) that not only support training on NVIDIA Jetson AGX and NANO devices, but also reduce the training time significantly compared to conventional DL architectures, rendering their feasibility for practical in-vehicle DMS deployment.

\section{Methodology}
\label{sec:methodology}
This section presents our proposed FedADAS framework for collaborative learning among AI-capable vehicles, defining the main algorithm that utilizes local soft label generation and corresponding KD. Then, we introduce the yawn recognition pipeline for DMS, tailored to in-vehicle edge deployment.

\begin{figure}[t]
     \centering
     \includegraphics[width=0.9\textwidth]{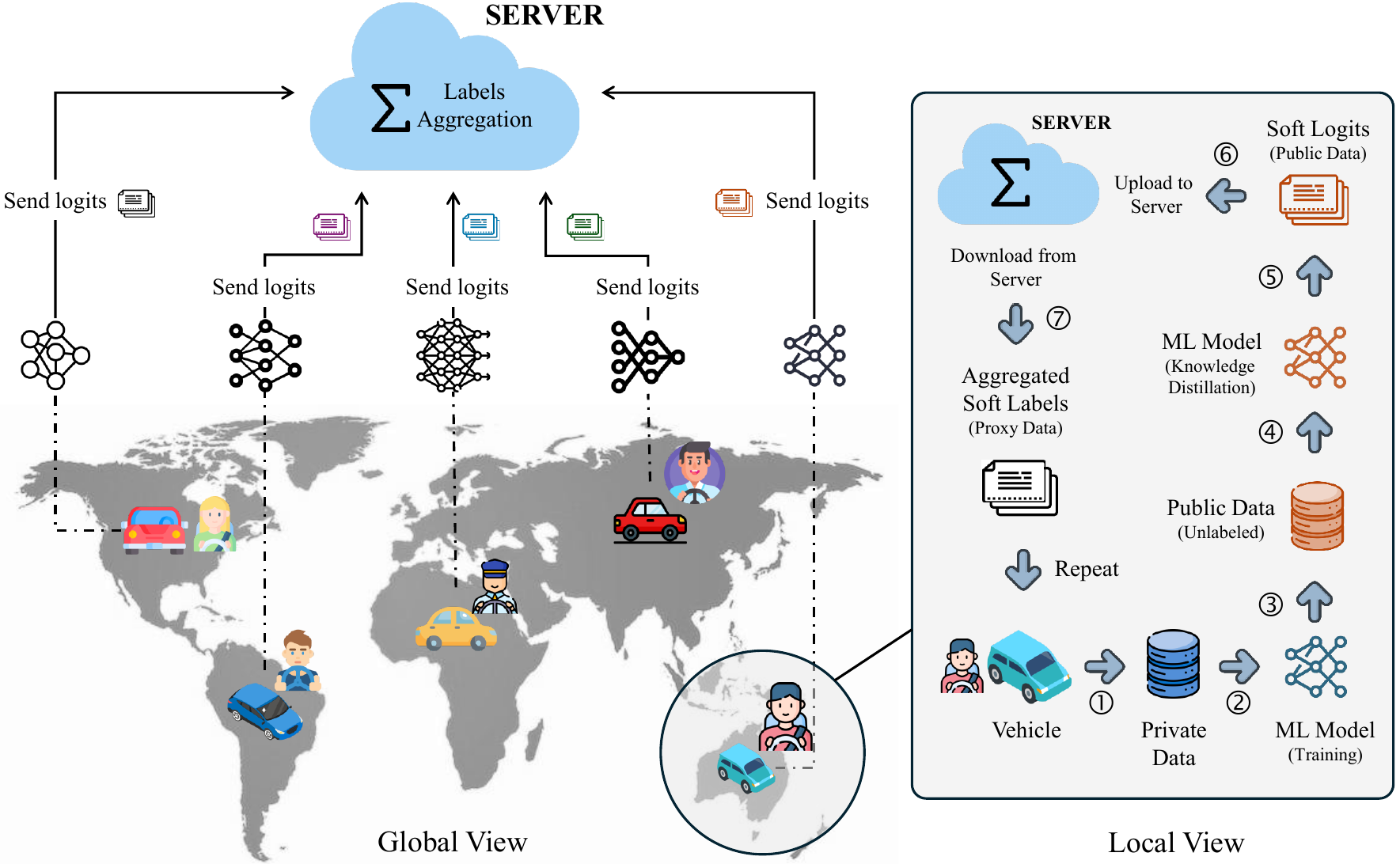}
     \caption{{FedADAS framework for collaborative learning in vehicular environments.} \textcolor{gray}{Icons from www.flaticon.com.}}
     \label{fig:FD}
\end{figure}

\subsection{Federated Distillation for Collaborative Learning Vehicles}
FedADAS enables cross-vehicular learning through KD without sharing full model parameters. It supports collaborative learning among vehicles equipped with custom edge devices, each running a distinct ML model architecture tuned for its specific resources. Fig.\ref{fig:FD} illustrates the FedADAS workflow at two scales.
\begin{itemize}
\item \emph{Global view:} Each vehicle processes local data with an on-board model, and transmits the resulting logits to a central server, which aggregates them into soft labels.
\item \emph{Local view:} The vehicle downloads the aggregated soft labels, pairs them with a shared public dataset, and performs KD to refine its own model before starting the next global cycle.
\end{itemize}

\noindent Algorithm \ref{alg:fd} presents the collaborative FD learning algorithm considering \enquote{$N$} geographically distributed vehicles holding local datasets $D_i$ = {($x_{i,j}$, $y_{i,j}$)} and training customized modes $M_i$. The algorithm proceeds in three phases that repeat for $T$ communication rounds.

\paragraph{Initialization (Lines 1-8):} Each vehicle locally initializes a customized DL model $\mathcal{M}_i$ according to its available resources. The objective is to train $\mathcal{M}_i$ to generalize across unseen data from other vehicles $i$. The server curates the unlabeled public dataset $\mathcal{D}_{\text{pub}}$, combined with open driving benchmarks and data gathered in controlled test runs, and shares it across all participating vehicles for KD.

\begin{algorithm}[t!]
\small
\caption{FD learning algorithm for collaborative vehicles}
\label{alg:fd}
\scalebox{0.7}{
\begin{minipage}{\linewidth}
\begin{algorithmic}[1]
\Require Local dataset $\{\mathcal{D}_i\}_{i=1}^N$ for $N$ clients (vehicles) each with $\{\mathcal{M}_i\}_{i=1}^N$ models, temperature parameter ($\tau$), number of communication rounds ($T$), local training epochs ($E_{local}$), distillation epochs ($E_{distill}$), and public dataset size ($S$).
\vspace{0.5em}
\State $\mathcal{D}_{pub} \leftarrow \emptyset$     \Comment{Initialization}
\For{each client \(i\!\in\!{N}\) \textbf{in parallel}}
    \State Initialize model $\mathcal{M}_i$ with random weights 
    \State $\mathcal{D}_{pub}^i \gets$ Get $S$ samples from $\mathcal{D}_i$
    \State \(\Call{SendToServer}{i,\mathcal{D}_{pub}^i}\)
\EndFor
\State \(\mathcal{D}_{\text{pub}} \gets \Call{ServerAggregate}{\{\mathcal{D}^i_{pub}\}_{i\in N}}\)
\State \(\Call{SendToClients}{N, \mathcal{D}_{pub}}\)
\vspace{0.5em}
\For{round $t = 1, 2, \ldots, T$}
    \State \(\mathcal{I}_r \gets \Call{ServerSelectRandomIndices}{\mathcal{D}_{pub}}\) 
    \For{each client \(i\!\in\!{N}\) \textbf{in parallel}} 
        \For{epoch \(e\!\in\!{E_{local}}\)} \Comment{Local Training}
                \State $\mathcal{M}_i \gets \Call{Train}{\mathcal{M}_{i}, \mathcal{D}_i}$
        \EndFor
        \State \(\mathcal{B} \gets \Call{ExtractSamples}{\mathcal{D}_{pub}, \mathcal{I}_r}\)
        \State \(y_\mathcal{B}^i \gets \Call{SoftmaxLogits}{\frac{\mathcal{M}_i(\mathcal{B})}{\tau}}\)
        \State \(\Call{SendToServer}{i,y_\mathcal{B}^i}\)
    \EndFor
    \State \(y_\mathcal{B} \gets \Call{ServerAggregate}{\{y_\mathcal{B}^i\}_{i\in N}}\)  \Comment{Ensemble Soft Labels}
    \State \(\Call{SendToClients}{N,y_\mathcal{B}}\)
    \vspace{0.5em}
    \For{each client \(i\!\in\!{N}\) \textbf{in parallel}} \Comment{Knowledge Distillation}
        \For{epoch \(e\!\in\!{E_{distill}}\)}
                \State \(y_{\mathcal{B}}^i \gets \Call{LogSoftmaxLogits}{\frac{\mathcal{M}_i(\mathcal{B})}{\tau}}\)
                \State $\mathcal{L}_{KD}^{i} = \tau^2 \cdot \text{KL}\left(y_{\mathcal{B}}^i \parallel y_\mathcal{B}\right)$
                \State \(\mathcal{M}_i \gets \Call{UpdateModel}{\mathcal{M}_i, \mathcal{L}_{KD}^{i}}\)
        \EndFor
    \EndFor
\EndFor
\end{algorithmic}
\end{minipage}
}
\end{algorithm}

\paragraph{Local training with soft label generation (lines 9-20):} 
For each communication round, the server selects random indices ($\mathcal{I}_r$) from $\mathcal{D}_{\text{pub}}$ and shares them with all vehicles for KD. Each vehicle first performs training on its local dataset $\mathcal{D}_i$ using cross-entropy loss for $E_{\text{local}}$ epochs using standard supervised learning. Then, it shares softmax logits on $\mathcal{B}$ with the server, where $\mathcal{B} \subset \mathcal{D}_{\text{pub}}$ based on $\mathcal{I}_r$. The server generates ensembled soft labels by averaging softmax logits and redistributes them back to the vehicles. 

\paragraph{Knowledge distillation (lines 21-28):} Each vehicle $i$ performs KD on $\mathcal{B}$ and minimizes the Kullback-Leibler (KL) divergence between the model's predictions ($y_{\mathcal{B}}^i$) and the ensemble soft labels ($y_\mathcal{B}$). Higher temperatures $\tau \gg 1$ produce softer probability distributions that expose relative similarities between classes, enabling transfer of structural knowledge about class relationships~\cite{hinton2015}. Line 23 applies $\log \text{softmax}$ to the student logits to ensure mathematical consistency in KL divergence computation. The temperature scaling factor $\tau^2$ in Line 24 compensates for gradient magnitude variations, as the gradients produced by soft targets scale as $1/\tau^2$~\cite{hinton2015}. This distillation process is performed for $E_{\text{distill}}$ epochs, enabling each model $\mathcal{M}_i$ to learn knowledge from the global ensemble logits ($y_\mathcal{B}$) while maintaining model heterogeneity. 

\begin{figure}[t]
     \centering
     \includegraphics[width=0.9\textwidth, trim=0cm 6.5cm 5.6cm 0cm, clip]{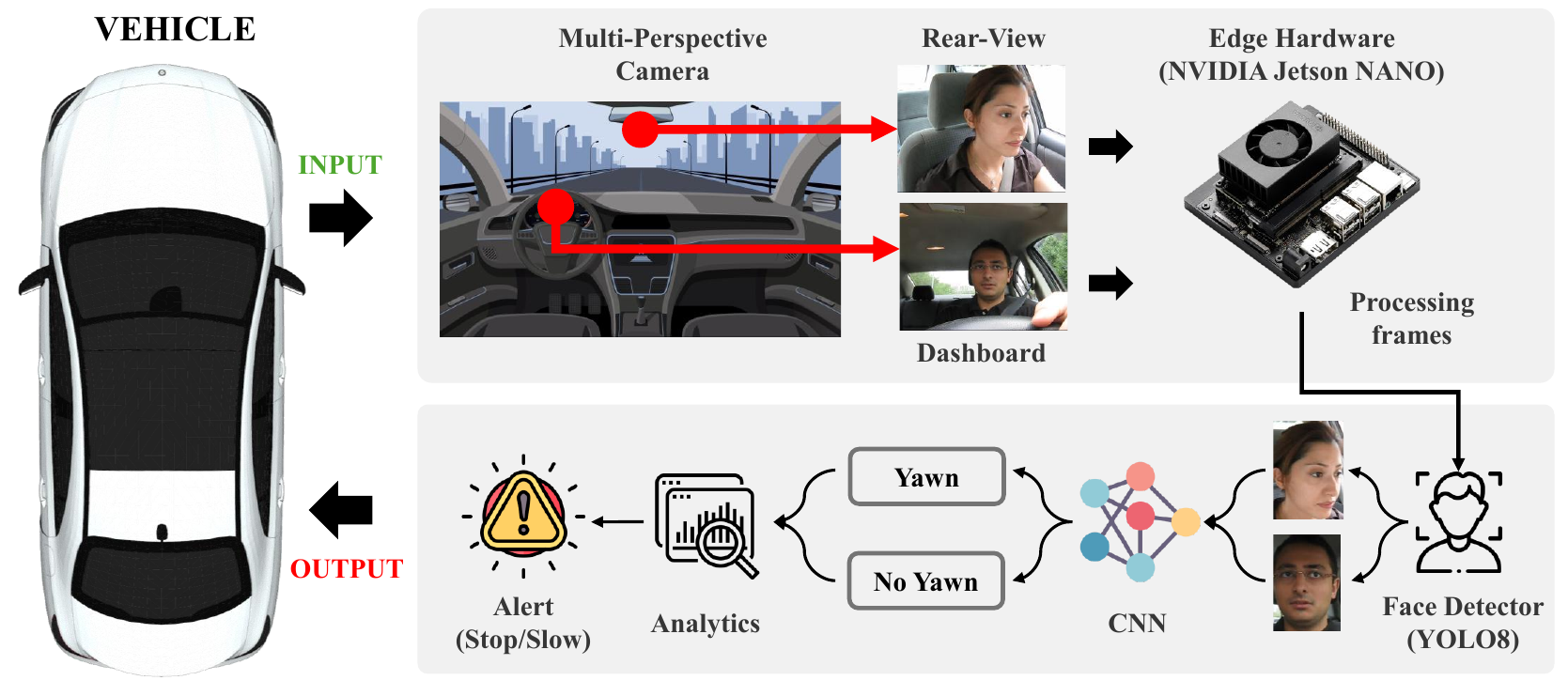}
     \caption{DMS pipeline for driver yawn recognition using DL models on edge devices. \textcolor{gray}{Icons from www.flaticon.com.}}
     \label{fig:yawn_pipeline}
\end{figure}

\subsection{DMS Pipeline for Driver Yawn Recognition}
Fig.~\ref{fig:yawn_pipeline} illustrates the deployment pipeline of the driver-yawn system for edge devices, covering both classification and detection models. The pipeline adopts a multi-stage architecture adapted for edge deployment.

\begin{enumerate}
\item \emph{Multi-perspective cameras} mounted on the dashboard and rear-view inside the vehicle's interior capture driver monitoring data further processed by the edge hardware (NVIDIA Jetson NANO).
\item \emph{Face detector} based on YOLO8~\cite{varghese2024yolov8} processes the captured frames from the camera feeds and localizes and extracts the driver's facial region. This step filters out irrelevant visual content, such as passengers, vehicle interior components, and external scenery, as it does not contribute to yawn detection and could interfere with model convergence. 
\item \emph{CNN} incorporate yawn classification and detection models that return two output labels on the cropped driver faces, \enquote{yawn} and \enquote{no yawn}, while the detection model adds \enquote{bounding boxes} localizing mouth regions. 
\item \emph{Analytics} module monitors yawn output frequency over time, enabling the generation of contextual alerts when drowsiness indicators exceed predetermined thresholds (\(FPS \times 2\)). Upon exceeding the threshold, the pipeline triggers vehicle safety interventions through \enquote{Stop/Slow} alerts.
\end{enumerate}

\noindent We introduce two edge-oriented yawn classification architectures\footnote[2]{Released open source at \href{https://opensource.silicon-austria.com/mujtabaa/fedadas/-/blob/main/assets/efficient_models.py}{\texttt{/fedadas/assets/efficient\_models.py}}}.
\begin{enumerate}
\item Memory-Efficient (\emph{ME-Net}) architecture incorporates lightweight depth-wise and point-wise layers to minimize memory footprint; 
\item Performance-Efficient (\emph{PE-Net}) architecture provides better performance at the cost of additional memory requirement.
\end{enumerate}
Both architectures support training and inference on edge hardware without requiring additional optimization techniques, such as quantization or pruning.

\begin{table}[t]
\centering
\caption{Performance comparison of conventional DL models with our proposed yawn classification models on NVIDIA A100, AGX, and NANO. The best results are marked in bold, and ($\dagger$) refers to our reproduced results from YawDD+~\cite{yawdd+}.}
\label{tab:yawn_cls_models_comparison}
\resizebox{\textwidth}{!}{%
\begin{tabular}{@{}lcccccccc@{}}
\toprule
\multirow{3}{*}{\textbf{\begin{tabular}[c]{@{}c@{}}Models\\ (Classification)\end{tabular}}} & \multirow{3}{*}{\textbf{\begin{tabular}[c]{@{}c@{}}Accuracy\\ (\%)\end{tabular}}} & \multirow{3}{*}{\textbf{\begin{tabular}[c]{@{}c@{}}F1-Score\\ (\%)\end{tabular}}} & \multirow{3}{*}{\textbf{\begin{tabular}[c]{@{}c@{}}Model Size\\ (MB)\end{tabular}}} & \multicolumn{3}{c}{\multirow{2}{*}{\textbf{\begin{tabular}[c]{@{}c@{}}Inference Time\\ (ms)\end{tabular}}}} & \multicolumn{2}{c}{\multirow{2}{*}{\textbf{\begin{tabular}[c]{@{}c@{}}Epoch Time\\ (min)\end{tabular}}}} \\ &  &  &  &  &  & \\ \cmidrule(l){5-9}
 &  &  &  & \textit{A100} & \textit{AGX} & \textit{NANO} & \textit{AGX} & \textit{NANO} \\ \midrule
DenseNet121                       & 99.11  & 97.47  & 27.1          & 18.34  & 33.46  & 48.62  & 24.68  & 48.72  \\
EfficientNet                    & 99.16  & 97.63  & 15.6          & 9.98   & 19     & 26.64  & 16.00  & 32.24  \\ 
EfficientNetv2                    & 99.02  & 97.21  & 77.8          & 20.79  & 38.59  & 56.37  & 27.31  & 58.40  \\ 
MNasNet$^\dagger$                           & 99.34  & 98.11  & 3.8           & 6.69   & 12.03  & 16.71  & 8.69   & 13.93  \\
MobileNetv2                       & 98.96  & 97.09  & 8.7           & 7      & 12.36  & 17.47  & 11.73  & 22.62  \\ 
MobileNetv3                       & 99.19  & 97.67  & 5.9           & 7.14   & 12.08  & 17.56  & 8.54   & 9.24   \\ 
RegNetx400mf                      & 99.21  & 97.75  & 19.7          & 10.44  & 19.08  & 26.16  & 12.37  & 17.93  \\ 
ResNet18                          & 99.27  & 97.92  & 42.7          & 3.50   & 6.09   & 8.43   & 8.82   & 16.97  \\ 
ResNeXt50                         & 98.99  & 97.13  & 88            & 8.18   & 14.14  & 19.94  & 24.28  & 58.33  \\ 
ShuffleNetv2                      & 99.31  & 98.03  & 1.4           & 8.80   & 19.7   & 14.09  & 9.08   & \textbf{7.39}  \\ 
SqueezeNet                       & 98.91  & 96.91  & 2.8           & 2.70   & 5.63   & 7.74   & 8.09   & 15.92  \\
Wide ResNet50                    & 99.03  & 97.27  & 255.3         & 8.21   & 14.26  & 20.60  & 31.35  & 75.25 \\ 
\textbf{ME-Net (Ours)}  & 99.30  & 97.99  & \textbf{0.6}  & 1.36   & 2.82   & 3.81   & \textbf{6.12}  & 9.53   \\ 
\textbf{PE-Net (Ours)}   & \textbf{99.39}   & \textbf{98.25}   & 99.7 & \textbf{1.31}  & \textbf{1.94}  & \textbf{1.99} & 7.96 & 14.45  \\ \midrule \midrule
\textbf{Model (Detection)}   & \textbf{mAP50-95 (\%)}   & \textbf{mAP50 (\%)}   &  &  & & & & \\ \midrule
YOLO11x          & 95.41           & \textbf{99.41}  & 114.4         & 16             & 44.80           & 153.1         & 308 & 777 \\
\textbf{YOLO11n$^\dagger$} & \textbf{95.69}  & 99.40           & \textbf{5.2}  & \textbf{9.5}   & \textbf{24.33}  & \textbf{35.7} & 41.4 & 85.8 \\ \bottomrule
\end{tabular}}
\end{table}

\section{Evaluation}
\label{sec:experiments}
We conducted extensive evaluation on yawn recognition for DMS (Sec.~\ref{subsec:exp_yawn_models}) and FedADAS (Sec.~\ref{subsec:exp_fd}) using the YawDD \cite{abtahi2014yawdd} and YawDD+ \cite{yawdd+} datasets.

\subsection{DMS Pipeline Evaluation}
\label{subsec:exp_yawn_models}
\textbf{Evaluation Process.}
All models were initialized with ImageNet pre-trained weights from PyTorch and adapted with a binary classification head (\enquote{yawn} and \enquote{no-yawn}). Input images were resized to $224\times224$ pixels and normalized using ImageNet statistics. Training data augmentation included random horizontal flips, random rotations ($\pm$10 degrees), and color jitter (brightness=0.2, contrast=0.2); validation and test sets were limited to resizing and normalization. Training used a batch size of 16, the Adam optimizer (lr = 0.001), and a reduce-on-plateau scheduler (patience = 3, factor = 0.1) with cross-entropy loss on both an A100 GPU and NVIDIA edge devices (AGX Orin and Jetson NANO). For yawn detection, we employed YOLO11~\cite{khanam2024yolov11} in both lightweight (n) and full-scale (x) configurations. YOLO11n was trained with a batch size of 8 on on AGX Orin and Jetson NANO, whereas YOLO11x required a reduced batch size of 2 on Jetson NANO due to memory constraints. Detection inputs were resized to $640\times640$ pixels.

\begin{table}[t]
\centering
\caption{Comparison with existing classification (Cls) and detection (Det) studies for yawn recognition using YawDD ($\dagger$) and YawDD+($\ddagger$).}
\label{tab:comparison}
\resizebox{\columnwidth}{!}{%
\begin{tabular}{@{}lccccc@{}}
\toprule
\textit{Approach}            & \textit{Method}    & \textit{Task}     & \textit{Accuracy/mAP}         & \textit{Edge Device} & \textit{Inference Time} \\ \midrule
\multirow{3}{*}{Video-based} & $\dagger$ 2s-STGCN~\cite{bai2021two} & Cls  & 93.4\%  & \NA  & \NA  \\
& $\dagger$ CNN-RNN~\cite{majeed2023detection} & Cls & 96.6\% & \NA  & \NA \\
& $\dagger$ DLS~\cite{xu2025novel} & Cls  & 96.14\%  & \NA & \NA\\  \midrule
\multirow{5}{*}{Frame-based} & $\dagger$ Two-stage CNN~\cite{he2020real}  & Cls & 93.83\%  & Raspberry Pi 4 & 96.3 ms \\ 
& $\dagger$ CNN~\cite{civik2023real} & Cls  & 94.5\% & NANO  & 166 ms \\ 
& $\dagger$ CNN, YOLO5, YOLO8~\cite{essahraui2025real} & Cls, Det, Det     & 93.31\%, 90.1\%, 90.3\%   & \NA & \NA  \\ 
& $\ddagger$ MNasNet, YOLO11~\cite{yawdd+}   & Cls, Det & 99.34\%, 95.69\% & NANO          & 16.71 ms, 35.7 ms  \\ 
\rowcolor{gray!50} & $\ddagger$ PE-Net, YOLO11x (Ours) & Cls, Det  & 99.39\%, 95.41\% & NANO & 1.99 ms, 153.1ms       \\ \bottomrule
\end{tabular}}
\end{table}

\textbf{Accuracy vs. Model Size.} Table~\ref{tab:yawn_cls_models_comparison} presents classification performance across all evaluated architectures, with accuracy consistently ranging from 98.91\% to 99.39\%. These results indicate that model size does not necessarily correlate with classification accuracy for the yawn recognition task. Among the proposed architectures, the \emph{PE-Net} model achieves the highest accuracy (99.39\%) and F1-score (98.25\%), surpassing all baselines and methods reported in Table~\ref{tab:comparison} with a model footprint of 99.7~MB. The \emph{ME-Net} prioritizes compactness, achieving a 166 $\times$ size reduction (0.6~MB memory footprint) while maintaining comparable accuracy (99.30\%). This performance is consistent with ShuffleNetV2~~\cite{shufflenetv2}, which achieves 99.31\% accuracy at 1.4~MB model footprint. For detection-based yawn recognition, YOLO11n achieves 95.69\% mAP50-95 while maintaining a compact 5.2~MB memory footprint, representing 22$\times$ size reduction compared to YOLO11x (114.4~MB) with minimal performance degradation. 

\textit{Key finding:} In case of yawn detection and classification, task-specific, light\-weight architectures consistently achieve competitive or superior accuracy compared to conventional DL models, while significantly reducing memory requirements and improving inference efficiency.

\textbf{Efficiency Analysis.} To evaluate deployment suitability on resource-cons\-trained platforms, we define two composite efficiency metrics ($\eta_{\text{inference}}$ and $\eta_{\text{training}}$) for the analyzed models:
\begin{equation}
\label{eq:inference_eff}
\eta_{\text{inference}} = \frac{\text{FPS} \times \text{accuracy}}{\text{model size}}
\end{equation}
\begin{equation}
\label{eq:training_eff}
\eta_{\text{training}} = \frac{\text{accuracy}}{\text{epoch training time} \times \text{model size}}
\end{equation}

These metrics capture the multidimensional trade-offs between accuracy, throughput, and resource consumption inherent to edge deployment and learning, enabling comparison beyond individual performance indicators. A model is considered inference efficient when it delivers high FPS, high accuracy, and a small memory footprint. Similarly, we consider that a model is training efficient when it delivers high accuracy, a small memory footprint, and a short epoch training time.

\begin{figure}[t]
     \centering
     \begin{subfigure}[b]{0.49\textwidth}
         \centering
         \includegraphics[width=\textwidth, trim=1.5cm 4.3cm 1.3cm 1.3cm, clip]{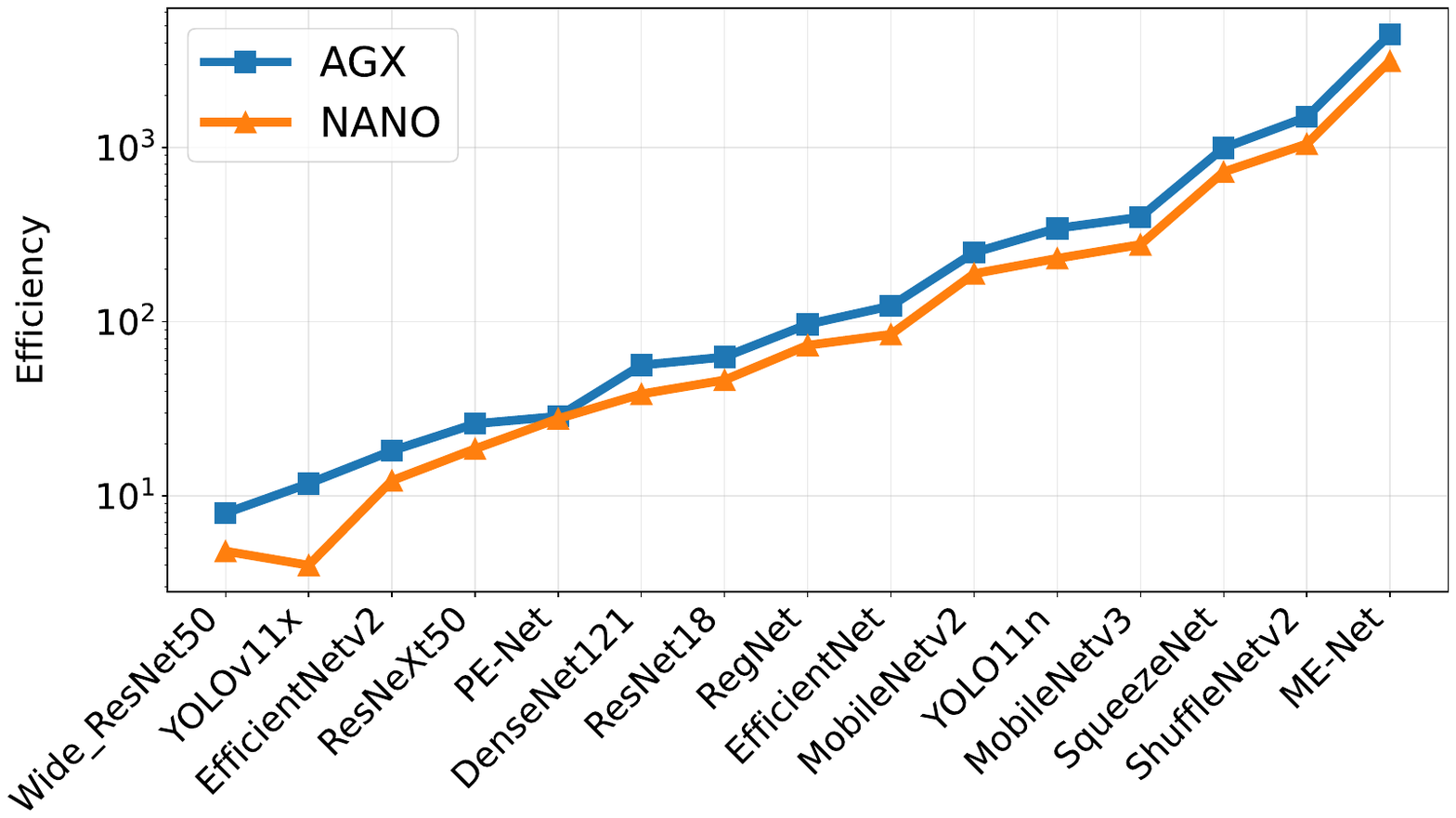}
         \caption{Inference efficiency - $\eta_{\text{inference}}$ (Eq.~\ref{eq:inference_eff})}
         \label{subfig:inf_efficiency}
     \end{subfigure}
     \hfill
     \begin{subfigure}[b]{0.49\textwidth}
         \centering
         \includegraphics[width=\textwidth, trim=1.1cm 4.3cm 1.3cm 1.3cm, clip]{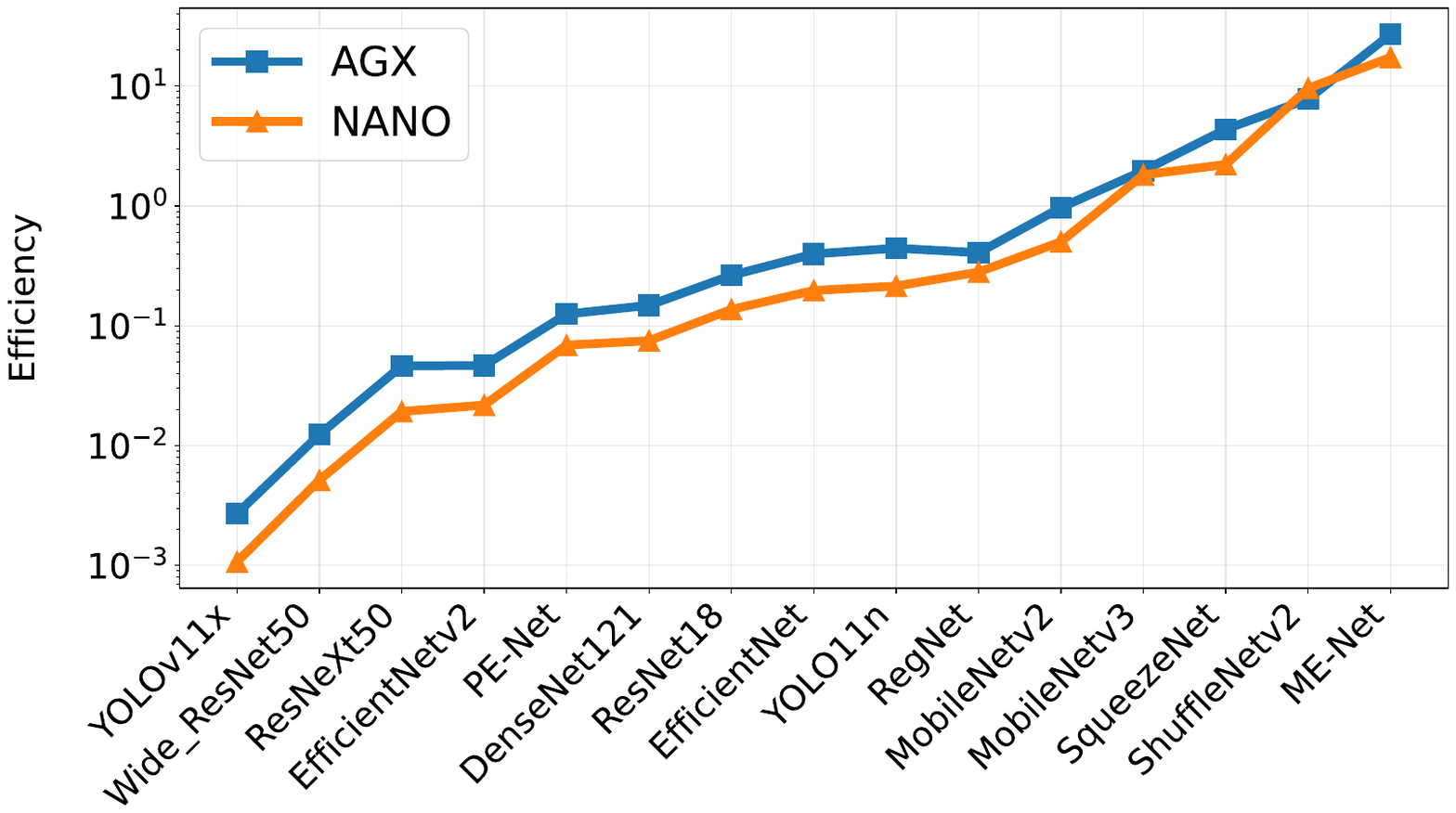}
         \caption{Training efficiency - $\eta_{\text{training}}$ (Eq.~\ref{eq:training_eff})}
         \label{subfig:train_efficiency}
     \end{subfigure}
    \caption{Training and inference efficiency scores of the DMS yawn recognition pipeline across various deep learning architectures.}
    \label{fig:efficiency}
\end{figure}

As shown in Fig.~\ref{fig:efficiency}, the \emph{ME-Net} model achieves the highest inference and training efficiency scores on both Jetson AGX Orin and Jetson NANO platforms. For training, this model requires only 6.12~min/epoch on AGX Orin and 9.53~min/epoch on Jetson NANO. In contrast, YOLO11x requires several hours per epoch even with reduced batch sizes, rendering on-device training infeasible for large-scale detection architectures. Although the \emph{PE-Net} model delivers the shortest inference time in all tests (see Table~\ref{tab:yawn_cls_models_comparison}), its overall inference and training efficiency are lower, and its training time is longer (7.96 min per epoch on AGX Orin). This difference reflects the architectural design: the \emph{PE-Net} network flattens the feature map to \num{50176} activations, inflating the fully connected layers and increasing memory traffic during backpropagation. By contrast, the \emph{ME-Net} design employs depth-wise separable convolutions and compresses the feature tensor to 256 channels with AdaptiveAvgPool, reducing both parameter count and gradient-update cost. These choices also shape inference characteristics. Standard 3$\times$3 convolutions are compute-bound and benefit from highly optimized GPU kernels, sustaining high throughput even in larger networks. Depth-wise separable convolutions and channel shuffling, as in ShuffleNetV2, lower theoretical FLOPs but become memory-bound and thus sensitive to memory latency. SqueezeNet follows a similar path, trimming parameters with fire modules while relying on standard convolutions to maintain GPU utilization.

\textit{Key finding:} Inference-optimized architectures do not necessarily provide training efficiency on edge devices, and raw inference speed alone is insufficient for evaluating deployment suitability. Composite metrics reveal that compact architectures utilizing efficient operations achieve superior overall efficiency for both inference and on-device training, despite potentially higher per-frame latency compared to larger, compute-optimized models.

\begin{table}[t]
\centering
\caption{Performance comparison between FedAvg and FedADAS under extreme non-IID conditions with different camera perspective.}
\label{tab:fedmd_results}
\resizebox{0.9\textwidth}{!}{%
\begin{tabular}{@{}ccccccc@{}}
\toprule
\textbf{\begin{tabular}[c]{@{}c@{}}Clients\\ (N)\end{tabular}} & \textbf{\begin{tabular}[c]{@{}c@{}}Camera\\ Perspective\end{tabular}} & \textbf{Frameworks}         & \textbf{\begin{tabular}[c]{@{}c@{}}Models\end{tabular}} & \textbf{\begin{tabular}[c]{@{}c@{}}Personalization\\ (Intra-Vehicle Accuracy)\end{tabular}} & \textbf{\begin{tabular}[c]{@{}c@{}}Generalization\\ (Inter-Vehicle Accuracy)\end{tabular}} & \textbf{BAM}         \\ \midrule
\multirow{4}{*}{3}      & \multirow{4}{*}{Dashboard}  & \multirow{2}{*}{\textbf{FedAvg}} & ME-Net   & 98.55    & 98.55   & 98.55                     \\
                        &                             &                         & \textbf{PE-Net}    & 99.78    & 99.78   & \textbf{99.78}                \\ 
\arrayrulecolor{lightgray}\cmidrule(l){3-7}\arrayrulecolor{black}
                        &                             & \multirow{2}{*}{FedADAS} & ME-Net   & 99.41    & 62.89   & 79.06                     \\
                        &                             &                         & PE-Net    & 99.50    & 99.15   & 99.33                \\ \midrule
\multirow{4}{*}{10}     & \multirow{4}{*}{Dashboard}  & \multirow{2}{*}{FedAvg} & ME-Net   & 97.83    & 97.83   & 97.83                     \\
                        &                             &                         & PE-Net    & 98.88    & 98.88   & 98.88                \\ 
\arrayrulecolor{lightgray}\cmidrule(l){3-7}\arrayrulecolor{black}
                        &                             & \multirow{2}{*}{\textbf{FedADAS}} & ME-Net   & 98.31    & 69.06   & 82.39                     \\
                        &                             &                         & \textbf{PE-Net}    & 99.78    & 98.09   & \textbf{98.93}                \\ \midrule
\multirow{4}{*}{25}     & \multirow{4}{*}{Dashboard}  & \multirow{2}{*}{FedAvg} & ME-Net   & 96.29    & 96.29   & 96.29                     \\
                        &                             &                         & PE-Net    & 95.95    & 95.95   & 95.95                \\ 
\arrayrulecolor{lightgray}\cmidrule(l){3-7}\arrayrulecolor{black}
                        &                             & \multirow{2}{*}{\textbf{FedADAS}} & ME-Net     & 98.46   & 83.97  & 90.65                     \\
                        &                             &                         & \textbf{PE-Net}    & 99.26    & 94.69   & \textbf{96.96}                \\ \midrule
\multirow{4}{*}{115}    & \multirow{4}{*}{Mirror-view}  & \multirow{2}{*}{FedAvg} & ME-Net   & 81.72    & 81.72   & 81.72                     \\
                        &                             &                         & PE-Net    & 76.35    & 76.35   & 76.35                \\ 
\arrayrulecolor{lightgray}\cmidrule(l){3-7}\arrayrulecolor{black}
                        &                             & \multirow{2}{*}{\textbf{FedADAS}} & ME-Net   & 97.18    & 67.64   & 81.07                     \\
                        &                             &                         & \textbf{PE-Net}    & 98.23    & 77.58   & \textbf{87.18}                \\ \bottomrule
\end{tabular}}
\end{table}

\subsection{FedADAS Evaluation}
\label{subsec:exp_fd}
\textbf{Evaluation Process.} The experimental setup considers $N$ participating non-IID vehicles, where $N\in[{3, 10, 25, 115}]$. From $N=3-25$, all drivers are dashboard-perspective; at $N=115$ we deliberately switch to mirror-view drivers to additionally introduce a covariate shift in camera geometry, simulating a worst-case fleet deployment. Each participating vehicle ($i\in N$) is assigned the data of a distinct driver, employs its own CNN architecture $\mathcal{M}_i$ and contributes nearly $10\%$ of its local data to a shared public dataset $\mathcal{D}_{pub}$ for KD. The procedure, implementing algorithm~\ref{alg:fd}, loops between local training and KD phases for $T=20$ communication rounds. KD is performed via Kullback-Leibler divergence minimization with temperature scaling parameter $\tau = 1.0$. Training hyperparameters include: batch size of $32$, learning rate of $0.001$ with Adam optimizer, and StepLR scheduler with $step\_size=10$ and $gamma=0.7$.

Table~\ref{tab:fedmd_results} presents an evaluation of performance of FedADAS and FedAvg under extreme non-IID conditions with varying client participation (3 to 115 clients) to assess real-world scalability. We report \emph{personalization} (accuracy on vehicle’s local data), \emph{generalization} (accuracy across vehicles), and their geometric mean, the Balanced Accuracy Metric (BAM), to capture the trade-off between these objectives. Table~\ref{tab:fedmd_results} reveals three key insights: 
\begin{itemize}
    \item FedADAS outperforms FedAvg once client diversity reaches ten vehicles, however demonstrates scale-dependent performance characteristics. With three clients the improvement is small, yet at $N=115$ the \emph{PE-Net} records a 21.88\% rise in personalization accuracy over FedAvg, confirming the robustness of soft-logit distillation under severe non-IID conditions.
    \item Cross-client knowledge transfer in lightweight models: \emph{ME-Net} generalization increased from 62.89\% at $N=3$ to 69.06\% at $N=10$ and 83.97\% at $N=25$, however at $N=115$, \emph{ME-Net} generalization reduces to 67.64\%. Because the server averages softmax logits across heterogeneous models, the soft labels ensemble is effectively pulled toward models producing higher-confidence (sharper) logits. Under domain shift, \emph{ME-Net} is regularized toward targets it lacks the capacity to represent, widening rather than closing the 9.94\% generalization gap and revealing capacity limitations under severe distribution shifts.
    \item Model capacity governs KD across clients. The persistent 10-15\% generalization gap between \emph{PE-Net} and \emph{ME-Net} models (10.72\% at $N=25$, 9.94\% at $N=115$) indicates that extreme capacity differences (166× parameters) limit KD effectiveness. This is consistent with Hinton et al.~\cite{hinton2015}, the KD transfer is bounded by student capacity where the knowledge encoded in inter-class similarity requires representational headroom that a 0.6 MB (\emph{ME-Net}) model cannot provide against a 99.7 MB (\emph{PE-Net}) model.
\end{itemize}

\textbf{Edge Latency and Communication Profiling.} Table~\ref{tab:profiling_FL_FD} summaries training time and communication costs for FedADAS and FedAvg across A100 GPU and two NVIDIA Jetson edge platforms (AGX Orin and Jetson NANO). By transmitting only soft logits (0.02 MB per round), FedADAS cuts uplink traffic by a factor of 60 relative to FedAvg with \emph{ME-Net} model (1.2 MB) and by nearly \num{10000} relative to FedAvg with \emph{PE-Net} model (199.4 MB). Local training time in FedADAS matches FedAvg, and the knowledge-distillation step adds negligible overhead: $0.20s-0.23s$ on A100, $0.39s-0.49s$ on AGX, and $0.86s-1.17s$ on NANO. The resulting per-round training time for FedADAS ranges from $1.55s-1.65s$ (A100), $4.81s-6.62s$ (AGX), to $10.15s-15.02s$ (NANO), demonstrating 4× and 9× slowdowns from A100 to AGX and NANO, respectively. Both architectures remain trainable on edge devices with \emph{ME-Net} completing training rounds in under 11 seconds even on NANO, validating the feasibility of on-device collaborative learning for practical vehicular deployments.

\section{Discussion \& Future Works}
\label{sec:discussion_and_analysis}
FedAvg~\cite{mcmahan2017communication} faces two main limitations in vehicular settings: it enforces a single model across clients regardless of each edge device’s capabilities, and it imposes heavy communication overhead by transmitting full model parameters. FedADAS mitigates both issues by allowing customized models on clients and exchanging only soft logits, which dramatically cuts communication costs and suits heterogeneous vehicular networks. Nevertheless, several challenges remain unaddressed:

\textbf{Privacy Leakage.} We adopt a semi-honest server with curious clients and assume communication channels are not adversarially compromised. Under this scenario, FedAvg and FedADAS expose qualitatively different attack risks motivated from prior works. FedAvg's gradient sharing enables white-box attacks that can reconstruct training data from periodic model updates~\cite{zhang2020secret}. FedADAS reduces this risk by transmitting only soft logits, denying the server access to model parameters. In black-box attacks~\cite{zhang2022ideal}, where adversaries query client models to reconstruct local samples, FedAvg produces more higher-fidelity reconstructions than FedADAS due to gradient informativeness versus dark knowledge (logits)~\cite{shao2024selective}. FedADAS relies on a shared public dataset, which introduces a privacy risk: whether the set is derived from client subsets, open repositories, or synthetic samples~\cite{qin2024knowledge}, a semi-honest server and inquisitive clients can analyze it to infer sensitive information. FedAvg is free from such attack by avoiding data sharing, however, a semi-honest server can still utilize global model to infer useful information about clients.

\textbf{Scalability \& Public Dataset Quality.} In Table~\ref{tab:fedmd_results}, the effective public dataset grows in diversity from dashboard view ($N=3 \rightarrow 25$), then shifts perspective (mirror-view) at $N=115$. We observe three regimes in Table 4: 
\begin{enumerate*}
\item increasing driver diversity at fixed perspective monotonically improves ME-Net generalization (62.89 $\rightarrow$ 69.06 $\rightarrow$ 83.97\%), indicating that dataset size correlates with cross-vehicle transfer when samples are in-distribution; 
\item the same scaling introduces a 21.88\% personalization gain over FedAvg for PE-Net at $N=115$, confirming robustness to class-imbalance-induced heterogeneity;
\item perspective bias at $N=115$ drops ME-Net generalization to 67.64\% despite a $>10\times$ larger $D_{pub}$, showing that representativeness dominates size.    
\end{enumerate*}
This empirically establishes that instead of size, the quality of $D_{pub}$ bounds distillation effectiveness, motivating representative sampling-protocols, server-side auditing requirements and security controls to lower privacy risk which we identify as future work.

\begin{table}[t]
\centering
\caption{Performance and resource consumption comparison of FL and FedADAS across different hardware platforms. Results are averaged over training rounds.}
\label{tab:profiling_FL_FD}
\resizebox{\textwidth}{!}{%
\begin{tabular}{lcccccc}
\hline
\multicolumn{1}{c}{\textbf{Method}} & \textbf{Models} & \textbf{\begin{tabular}[c]{@{}c@{}}Communication Cost\\ Per Round\end{tabular}} & \textbf{Component} & \textbf{A100 (s)} & \textbf{AGX (s)} & \textbf{NANO (s)} \\ \hline
\multicolumn{1}{c}{\multirow{2}{*}{FedAvg}} & \textbf{ME-Net} & 1.2 MB & \multirow{2}{*}{Local Training} & \textbf{1.44 $\pm$ 0.11} & \textbf{5.53 ± 0.35s} & \textbf{9.74 ± 0.12s} \\
\multicolumn{1}{c}{} & PE-Net & 199.4 MB &  & 1.67 $\pm$ 0.03 & 6.95 ± 0.19s & 14.26 ± 0.11s \\ \hline
\multirow{4}{*}{FedADAS} & \multirow{2}{*}{\textbf{ME-Net}} & \multirow{4}{*}{0.02 MB} & Local Training & \textbf{1.35 $\pm$ 0.16} & \textbf{4.42 ± 0.15s} & \textbf{9.29 ± 0.12s} \\
 &  &  & \multicolumn{1}{c}{KD} & \textbf{0.20 $\pm$ 0.01} & \textbf{0.39 ± 0.14s} & \textbf{0.86 ± 0.07s} \\ \cline{2-2} \cline{4-7} 
 & \multirow{2}{*}{PE-Net} &  & Local Training & 1.42 $\pm$ 0.11 & 6.13 ± 0.10s & 13.85 ± 0.53s \\
 &  &  & \multicolumn{1}{c}{KD} & 0.23 $\pm$ 0.01 & 0.49 ± 0.06s & 1.17 ± 0.24s \\ \hline
\end{tabular}
}
\end{table}

\section{Conclusion}
\label{sec:conclusion}
This paper presents FedADAS, a first FD framework addressing critical challenges of traditional FL in vehicular networks, such as device heterogeneity, high communication overhead, and non-IID data distribution. This design cuts communication by 9974$\times$ relative to FL and maintains a strong personalization–generalization balance at higher client participation. The underlying logit-exchange and KL-divergence aggregation in FedADAS make no task-specific assumptions and applicable to multi-class perception tasks where the per-round communication cost scales linearly with the number of classes. The accompanying edge-oriented yawn classifiers reach 99.39\% accuracy with 1.99ms inference latency on Jetson NANO, supporting full on-device training. Extensive experiments confirm FedADAS’s scalability ($N=115$) and robustness in extreme non-IID settings, establishing a viable foundation for adaptive driver-monitoring across heterogeneous vehicle fleets. 

\subsubsection*{Acknowledgements.} This work received funding from the European Union MSCA COFUND project CRYSTALLINE (grant agreement 101126571), and the Austrian Research Promotion Agency (FFG grant agreement 909989 ``AIM AT Stiftungsprofessur für Edge AI''). This work has been supported by Silicon Austria Labs (SAL) owned by the Republic of Austria, the Styrian Business Promotion Agency (SFG), the federal state of Carinthia, the Upper Austrian Research (UAR), and the Austrian Association for the Electric and Electronics Industry (FEEI). This research was partially funded by the Austrian Science Fund (FWF) 10.55776/COE12.

%
%
%
\bibliographystyle{splncs04}
\bibliography{mybibliography}

@inproceedings{abtahi2014yawdd,
  title={YawDD: A yawning detection dataset},
  author={Abtahi, Shabnam and Omidyeganeh, Mona and Shirmohammadi, Shervin and Hariri, Behnoosh},
  booktitle={Proceedings of the 5th ACM multimedia systems conference},
  pages={24--28},
  year={2014}
}

@article{khanam2024yolov11,
  title={Yolov11: An overview of the key architectural enhancements},
  author={Khanam, Rahima and Hussain, Muhammad},
  journal={arXiv preprint arXiv:2410.17725},
  year={2024}
}

@misc{dms,
      title={Smart Driver Monitoring Robotic System to Enhance Road Safety : A Comprehensive Review}, 
      author={Farhin Farhad Riya and Shahinul Hoque and Xiaopeng Zhao and Jinyuan Stella Sun},
      year={2024},
      eprint={2401.15762},
      archivePrefix={arXiv},
      primaryClass={cs.RO},
}

@online{NVIDIA_FL_AV,
  author = {Kuo, Jerry and Wang, Xiaokang and Li, Xiaolong and Xu, Tianyu},
  title = {Federated Learning in Autonomous Vehicles Using Cross-Border Training},
  year = {2023},
  url = {https://developer.nvidia.com/blog/federated-learning-in-autonomous-vehicles-using-cross-border-training/},
  organization = {NVIDIA Developer},
  urldate = {2025-12-09}
}

@inproceedings{yawdd+,
      title={YawDD+: Frame-level Annotations for Accurate Yawn Prediction}, 
      author={Ahmed Mujtaba and Gleb Radchenko and Marc Masana and Radu Prodan},
      booktitle={Proceedings of the IEEE International Conference on Image Processing},
      year={2026},
}

@article{wang2025empowering,
  title={Empowering edge intelligence: A comprehensive survey on on-device ai models},
  author={Wang, Xubin and Tang, Zhiqing and Guo, Jianxiong and Meng, Tianhui and Wang, Chenhao and Wang, Tian and Jia, Weijia},
  journal={ACM Computing Surveys},
  volume={57},
  number={9},
  pages={1--39},
  year={2025},
  publisher={ACM New York, NY}
}

@misc{nthsa,
    author = {National Highway Traffic Safety Administration},
    title = {Drowsy Driving},
    publisher = {NTHSA},
    url = {https://www.nhtsa.gov/risky-driving/drowsy-driving}
}

@inproceedings{zhang2022ideal,
  title={Ideal: Query-efficient data-free learning from black-box models},
  author={Zhang, Jie and Chen, Chen and Lyu, Lingjuan},
  booktitle={International Conference on Learning Representations},
  year={2023}
}

@inproceedings{zhang2020secret,
  title={The secret revealer: Generative model-inversion attacks against deep neural networks},
  author={Zhang, Yuheng and Jia, Ruoxi and Pei, Hengzhi and Wang, Wenxiao and Li, Bo and Song, Dawn},
  booktitle={Proc. of the IEEE/CVF conference on computer vision and pattern recognition},
  pages={253--261},
  year={2020}
}

@inproceedings{varghese2024yolov8,
  title={Yolov8: A novel object detection algorithm with enhanced performance and robustness},
  author={Varghese, Rejin and Sambath, M},
  booktitle={2024 International conference on advances in data engineering and intelligent computing systems (ADICS)},
  pages={1--6},
  year={2024},
  organization={IEEE}
}

@article{shao2024selective,
  title={Selective knowledge sharing for privacy-preserving federated distillation without a good teacher},
  author={Shao, Jiawei and Wu, Fangzhao and Zhang, Jun},
  journal={Nature Communications},
  volume={15},
  number={1},
  pages={349},
  year={2024}
}

@inproceedings{FedMD,
  title={Fedmd: Heterogenous federated learning via model distillation},
  author={Li, Daliang and Wang, Junpu},
  booktitle={Proceedings of Neural Information Processing Systems, FLDPC Workshop},
  year={2019}
}

@article{qin2024knowledge,
  title={Knowledge distillation in federated learning: A survey on long lasting challenges and new solutions},
  author={Qin, Laiqiao and Zhu, Tianqing and Zhou, Wanlei and Yu, Philip S},
  journal={International Journal of Intelligent Systems},
  volume={2025},
  number={1},
  pages={7406934},
  year={2025},
  publisher={Wiley Online Library}
}

@inproceedings{hinton2015,
  title={Distilling the Knowledge in a Neural Network},
  author={Geoffrey Hinton and Oriol Vinyals and Jeff Dean},
  booktitle={Proceedings of Neural Information Processing Systems Workshop},
  year={2014}
}

@inproceedings{shufflenetv2,
  title={Shufflenet v2: Practical guidelines for efficient cnn architecture design},
  author={Ma, Ningning and Zhang, Xiangyu and Zheng, Hai-Tao and Sun, Jian},
  booktitle={Proceedings of the European conference on computer vision (ECCV)},
  pages={116--131},
  year={2018}
}

@article{he2020real,
  title={A real-time driver fatigue detection method based on two-stage convolutional neural network},
  author={He, Hu and Zhang, Xiaoyong and Jiang, Fu and Wang, Chenglong and Yang, Yingze and Liu, Weirong and Peng, Jun},
  journal={IFAC-PapersOnLine},
  volume={53},
  number={2},
  pages={15374--15379},
  year={2020},
  publisher={Elsevier}
}

@article{civik2023real,
  title={Real-time driver fatigue detection system with deep learning on a low-cost embedded system},
  author={Civik, Esra and Yuzgec, Ugur},
  journal={Microprocessors and Microsystems},
  volume={99},
  year={2023},
  publisher={Elsevier}
}

@article{majeed2023detection,
  title={Detection of drowsiness among drivers using novel deep convolutional neural network model},
  author={Majeed, Fiaz and Shafique, Umair and Safran, Mejdl and Alfarhood, Sultan and Ashraf, Imran},
  journal={Sensors},
  volume={23},
  number={21},
  pages={8741},
  year={2023},
  publisher={MDPI}
}

@article{bai2021two,
  title={Two-stream spatial--temporal graph convolutional networks for driver drowsiness detection},
  author={Bai, Jing and Yu, Wentao and Xiao, Zhu and Havyarimana, Vincent and Regan, Amelia C and Jiang, Hongbo and Jiao, Licheng},
  journal={IEEE Transactions on Cybernetics},
  volume={52},
  number={12},
  pages={13821--13833},
  year={2021},
  publisher={IEEE}
}

@article{xu2025novel,
  title={A Novel Driver Fatigue Detection Method Based on Dual-Stream Swin-Transformer},
  author={XU, Mingyang and ZHAN, Ao and WU, Chengyu and WANG, Zhengqiang},
  journal={IEICE Transactions on Information and Systems},
  pages={2024EDL8094},
  year={2025},
  publisher={The Institute of Electronics, Information and Communication Engineers}
}

@article{essahraui2025real,
  title={Real-time driver drowsiness detection using facial analysis and machine learning techniques},
  author={Essahraui, Siham and Lamaakal, Ismail and El Hamly, Ikhlas and Maleh, Yassine and Ouahbi, Ibrahim and El Makkaoui, Khalid and Filali Bouami, Mouncef and P{\l}awiak, Pawe{\l} and Alfarraj, Osama and Abd El-Latif, Ahmed A},
  journal={Sensors},
  volume={25},
  pages={812},
  year={2025},
  publisher={MDPI}
}

@article{al2025real,
  title={Real-time distracted driving detection based on gm-yolov8 on embedded systems},
  author={Al-Mahbashi, Mohammed and Li, Gang and Peng, Yaxue and Al-Soswa, Mohammed and Debsi, Ali},
  journal={Journal of Transportation Engineering, Part A: Systems},
  volume={151},
  number={3},
  pages={04024126},
  year={2025},
  publisher={American Society of Civil Engineers}
}

@inproceedings{zhou2021real,
  title={A real-time driver fatigue monitoring system based on lightweight convolutional neural network},
  author={Zhou, Chunyu and Li, Jun},
  booktitle={2021 33rd Chinese Control and Decision Conference (CCDC)},
  pages={1548--1553},
  year={2021},
  organization={IEEE}
}

@inproceedings{wang2024federated,
  title={Federated Learning with Knowledge Distillation to Mitigate Catastrophic Forgetting and Data Heterogeneity in IoV Systems},
  author={Wang, Jiayu and Gao, Jiechao},
  booktitle={2024 IEEE International Conference on Big Data (BigData)},
  pages={2914--2923},
  year={2024},
  organization={IEEE}
}

@article{eid2024federated,
  title={Federated learning system on autonomous vehicles for lane segmentation},
  author={Eid Kishawy, Mohab M and Abd El-Hafez, Mohamed T and Yousri, Retaj and Darweesh, M Saeed},
  journal={Scientific Reports},
  volume={14},
  number={1},
  pages={25029},
  year={2024},
  publisher={Nature Publishing Group UK London}
}

@article{liao2025personalized,
  title={Personalized federated learning through self-knowledge distillation in vehicular edge computing},
  author={Liao, Gengjian and Yang, Yulin and Feng, Zhenni},
  journal={Computer Networks},
  year={2025},
  publisher={Elsevier}
}

@article{huang2025environment,
  title={Environment-Aware Personalized Heterogeneous Federated Distillation for Dual-Layer Blockchain-Enabled Internet of Vehicles},
  author={Huang, Xiaoge and Li, Wenjing and Liang, Chengchao and Cao, Bin and Zhou, Mu},
  journal={IEEE Transactions on Vehicular Technology},
  year={2025},
  publisher={IEEE}
}

@article{xiao2025feddld,
  title={FedDLD: Dual-Level Federated Distillation with Adaptive Knowledge Transfer for DAG-secured IoVs},
  author={Xiao, Sa and Huang, Xiaoge and Zhou, Mu and Liang, Chengchao and Chen, Qianbin},
  journal={IEEE Transactions on Vehicular Technology},
  year={2025},
  publisher={IEEE}
}

@inproceedings{data-free,
  title={Federated Learning with Data-Free Distillation for Heterogeneity-Aware Autonomous Driving},
  author={Liang, Junyao and Li, Juan and Zhang, Ji and Zang, Tianzi},
  booktitle={2024 International Joint Conference on Neural Networks (IJCNN)},
  pages={1--7},
  year={2024},
  organization={IEEE}
}

@article{bano2024fedcmd,
  title={FedCMD: A federated cross-modal knowledge distillation for drivers’ emotion recognition},
  author={Bano, Saira and Tonellotto, Nicola and Cassar{\`a}, Pietro and Gotta, Alberto},
  journal={ACM Transactions on Intelligent Systems and Technology},
  volume={15},
  number={3},
  pages={1--27},
  year={2024},
  publisher={ACM New York, NY, USA}
}

@article{shang2023fedbikd,
  title={FedBiKD: Federated bidirectional knowledge distillation for distracted driving detection},
  author={Shang, Ertong and Liu, Hui and Yang, Zhuo and Du, Junzhao and Ge, Yiming},
  journal={IEEE Internet of Things Journal},
  volume={10},
  number={13},
  pages={11643--11654},
  year={2023},
  publisher={IEEE}
}

@inproceedings{mcmahan2017communication,
  title={Communication-efficient learning of deep networks from decentralized data},
  author={McMahan, Brendan and Moore, Eider and Ramage, Daniel and Hampson, Seth and y Arcas, Blaise Aguera},
  booktitle={Artificial intelligence and statistics},
  pages={1273--1282},
  year={2017},
  organization={PMLR}
}

@article{lan2023communication,
  title={Communication-efficient federated learning for resource-constrained edge devices},
  author={Lan, Guangchen and Liu, Xiao-Yang and Zhang, Yijing and Wang, Xiaodong},
  journal={IEEE Transactions on Machine Learning in Communications and Networking},
  volume={1},
  pages={210--224},
  year={2023},
  publisher={IEEE}
}

@inproceedings{mujtaba2025federated,
  title={Federated Distillation on Edge Devices: Efficient Client-Side Filtering for Non-IID Data},
  author={Mujtaba, Ahmed and Radchenko, Gleb and Prodan, Radu and Masana, Marc},
  booktitle={2025 3rd International Conference on Federated Learning Technologies and Applications (FLTA)},
  pages={228--235},
  year={2025},
}

\end{document}